\documentclass[preprints,article,accept,pdftex,moreauthors]{Definitions/mdpi} 

\usepackage{xcolor}
\usepackage{natbib}
\usepackage{amsmath}
\usepackage{amsfonts}
\usepackage{graphics}
\usepackage{graphicx}

\def\edot{$\dot{E}$}
\def\mdot{\ifmmode \dot M \else $\dot M$\fi}
\def\edot{\ifmmode \dot E \else $\dot E$\fi}
\def\tc{\tao_{c}}
\def\tc{\ifmmode \tau_{c} \else $\tau_{c}$\fi}
\def\k1{\ifmmode K_{1} \else $K_{1}$ \fi}
\def\k2{\ifmmode K_{2} \else $K_{2}$ \fi}
\newcommand{\Odot}{\dot{\Omega}}
\newcommand{\Oddot}{\ddot{\Omega}}
\newcommand{\Pdot}{\dot{P}}

\newcommand{\Rmnum}[1]{\uppercase\expandafter{\romannumeral #1}}

\def\apjl{Astrophys.~J.~Lett.}

\def\apj{Astrophys.~J.}

\def\mnras{Mon.~Not.~R.~Astron.~Soc.}
\def\aap{Astron.~Astrophys.}

\def\nat{Nature}
\def\aj{Astron.~J.}
\def\apss{Astrophys. Space Sci.}
\def\araa{Annu. Rev. Astron. Astrophys.}
\def\baas{Bull. Am. Astron. Soc.}

\firstpage{1}
\pubvolume{1}
\issuenum{1}
\articlenumber{0}
\pubyear{2022}
\copyrightyear{2022}
\externaleditor{Academic Editor: Firstname Lastname}
\datereceived{17 October 2022}
\dateaccepted{23 November 2022}
\datepublished{}
\hreflink{https://doi.org/} % If needed use \linebreak

\Title{Evolution of Spin Period and Magnetic Field of the Crab Pulsar: Decay of the  Braking Index by the Particle Wind Flow Torque}

\TitleCitation{Evolution of Spin Period and Magnetic Field of the Crab Pulsar: Decay of the  Braking Index by the Particle Wind Flow Torque}

\Author{{Cheng-Min Zhang} %MDPI: Please carefully check the accuracy of names and affiliations.
 $^{1,2,3,}${*}%MDPI: Newly added correspondence information. Please confirm. re: conform
$^{,\dagger}$\orcidA{}, {Xiang-Han Cui} $^{1,3}$\orcidB{}, Di Li $^{1,3,4,5}$\orcidC{}, {De-Hua Wang} %MDPI: The name of this author is different from the submitting system. Please confirm which one is correct.
 $^{6}$, {Shuang-Qiang Wang} $^{7}$, Na Wang $^{7}$, {Jian-Wei Zhang} $^{1}$, Bo Peng $^{1,3}$, {Wei-Wei Zhu} $^{1,3}$, {Yi-Yan Yang} $^{8}$ and {Yuan-Yue Pan} $^{9}$}

\AuthorNames{Cheng-Min Zhang, Xiang-Han Cui, Di Li, De-Hua Wang, Shuang-Qiang Wang, Na Wang, Jian-Wei Zhang, Bo Peng, Wei-Wei Zhu, Yi-Yan Yang, Yuan-Yue Pan}

\AuthorCitation{Zhang, C.-M.; Cui, X.-H.; Li, D.; Wang, D.-H.; Wang, S.-Q.; Wang, N.; Zhang, J.-W.; Peng, B.; Zhu, W.-W.; Yang, Y.-Y.; et al.}

\address{%
$^{1}$ \quad CAS Key Lab of FAST, National Astronomical Observatories, Chinese Academy of Sciences, \mbox{Beijing 100101, China}; {cuixianghan@nao.cas.cn (X.-H.C.); dili@nao.cas.cn (D.L.); jwzhang@bao.ac.cn (J.-W.Z.); pb@nao.cas.cn (P.B.); zhuww@nao.cas.cn (W.-W.Z.)} %MDPI: We added the email addresses here according to the submitting system. Please confirm.
\\
$^{2}$ \quad School of Physical Sciences, University of Chinese Academy of Sciences, Beijing 100049, China\\
$^{3}$ \quad School of Astronomy and Space Science, University of Chinese Academy of Sciences, Beijing 100049, China\\
$^{4}$ \quad NAOC-UKZN Computational Astrophysics Centre, University of KwaZulu-Natal, Durban 4000, South Africa\\
$^{5}$ \quad Research Center for Intelligent Computing Platforms, Zhejiang Laboratory, Hangzhou 311100, China\\
$^{6}$ \quad School of Physics and Electronic Science, Guizhou Normal University, Guiyang 550001, China; wangdh@gznu.edu.cn\\
$^7$ \quad Xinjiang Astronomical Observatory, Chinese Academy of Sciences, {Urumqi} %MDPI: we deleted Xinjiang here, because province information is not allowed here, please confirm
  830011, China; {wangshuangqiang@xao.ac.cn (S.-Q.W.); na.wang@uao.ac.cn (N.W.)}\\
$^{8} $ \quad School of Physics and Electronic Science, Guizhou Education University, Guiyang 550018, China; {yangyiyan@gznc.edu.cn}\\
$^{9} $ \quad School of Physics, Xiangtan University, {Xiangtan} 533001, China; {panyy@xtu.edu.cn}
}

\corres{Correspondence: zhangcm@bao.ac.cn}

\abstract{The evolutions of a neutron star's rotation and magnetic field (B-field) have remained unsolved puzzles for over half a century.
We ascribe the rotational braking torques of pulsar to both components, the standard magnetic dipole radiation (MDR) and particle wind flow ( MDR + Wind, hereafter named MDRW), which we apply to the Crab pulsar (B0531 + 21), the only source with a known age and long-term continuous monitoring by radio telescope.
Based on the above presumed simple spin-down torques, we obtain the exact analytic solution on the rotation evolution of the Crab pulsar,  together with the related outcomes as described below:
(1) unlike the constant characteristic  B-field suggested by the MDR model, this value for the Crab pulsar increases  by a hundred times in 50~kyr while its real B-field has no change;
(2) the rotational braking index evolves from $\sim$3 to 1 in the long-term, however, it drops from 2.51 to 2.50 in $\sim$45 years at the present stage,
while the particle flow contributes approximately 25\% of the total rotational energy loss rate;
(3) strikingly, the characteristic age has the maximum limit of $\sim$10 kyr, meaning that it  is not always a good indicator of a real age.
Furthermore, we discussed the evolutionary path of the Crab pulsar from the MDR to the wind domination by comparing with the possible wind braking candidate pulsar PSR J1734-3333.
}


\begin{document}

\section{Introduction} \label{1}

About 55 years has passed since the first pulsar was discovered in 1967 \cite{Hewish68}, identified as the beacon phenomena of rotating neutron stars (NSs) \cite{Pacini68, Gold68}.
{ From then on, more than 3500 radio pulsars have been observed \cite{Manchester05}, including over 600 ones recently detected by five-hundred-meter aperture spherical radio telescope (FAST) \cite{Li18, Han21, Wang21, Miao22}, but the puzzle of how the rotation and magnetic field of pulsar evolves remains \cite{Lorimer19, Philippov22}.}
To answer these fundamental questions, the Crab pulsar (PSR B0531 + 21) is usually considered one of the best astronomical labs because it is the only pulsar with a known age.
{ This famous pulsar was discovered in 1968 in the Crab nebula with a supernova remnant (SNR) \cite{Lovelace68, Staelin68,Lovelace12,Alsabti17} that was identified as a product of a massive star explosion in 1054 AD with a clear historical record by ancient Chinese astronomers \cite{Manchester77,Lyne12}.}
In addition, this young pulsar has been continuously monitored for half a century, yielding fruitful and accurate observational data~\cite{Lyne93, Lyne15}.

Recently, the magnetic dipole radiation (MDR) model \cite{Gunn69, Pacini68} was proposed and developed to account for the rotation slowdown of pulsar, in which the loss rate of the rotational kinetic energy ($\dot{E}$) of pulsar is supposed to equate the emission power of a magnetic dipole in vacuum $L_{d}= K_{1} \Omega^{4}$.
{ For example, }a perpendicular rotator with $\dot{E}\equiv L_{d}$ infers that $-I\Omega\Odot = K_{1}\Omega^{4}$, where the definitions of the conventional quantities are $K_{1}=2B^2R^6/3c^3$; I ($10^{45}$ g cm$^{2}$)---the moment of inertia; $\Omega$ ($\Odot$)---the rotation angular frequency (its derivative); $B$---the NS surface magnetic field (B-field); R ($10^{6}$ cm)---the stelar radius; and $c$---the speed of light \cite{Manchester77,Lyne12,Lorimer12}. 
This standard MDR model has successfully predicted the B-field strength of the Crab pulsar and normal pulsars by the observed timing quantities { as follows}.
{ In terms of} spin period ($P=2\pi/\Omega$) and its derivative $\Pdot$, the derived characteristic B-field is $B = B_{ch} \equiv [{3c^{3}I/(8\pi^{2}R^{6}})]^{1/2} (P \Pdot)^{1/2}\simeq 10^{12} (P \Pdot/10^{-15})^{1/2} G $ \cite{Manchester77,Lyne12, Lorimer12}, which is very close to the direct measurement by the cyclotron absorption lines of X-rays for an accreting NS in the high mass X-ray binary (HMXB) \cite{Truemper78,Bhattacharya91}.
Moreover, the evolution of the NS characteristic B-field has been widely investigated by previous studies \cite{Narayan90,Lorimer97,Geppert02,Faucher06}; and for a recent review, we refer to Ref.~\cite{Igoshev21}.

Without the direct measurement of B-field for a normal pulsar,  the validity of the characteristic B-field as a replacement of the real B-field is often debated.
To answer above doubts, astronomers can measure the braking index ($n$) of a pulsar that is defined by the spin timing parameters as
\begin{equation}
n\equiv\Omega\Oddot/\Odot^2\;,
\label{index}
\end{equation}
where $\Oddot$ is the second derivative of angular frequency, whereby the theoretical canonical constant value of \emph{n} = 3 is obtained for the MDR model \cite{Manchester85,Blandford88}.
However, for the Crab pulsar, \emph{n} = 2.515 was first reported in the 1970s \cite{Boynton72,Groth75}, and its continuous monitoring followed the stable and accurate values of \emph{n} = 2.51 in 1993 \cite{Lyne93} and \emph{n} = 2.50 in 2015 into the present, constituting an effort of 45 years \cite{Lyne15}. 
Owing to the fact that the accurate measurement of braking, the index %English editor: Please ensure that the meaning has been retained.
requires pulsars with
a high-$\dot{E}$ or high-$\dot{P}$.
{ In general}, only eight young radio pulsars have been measured, and  to date, the stable values of braking indices have approximately ranged from 1 to 3 \cite{Lyne15}, which is deviated from the assumption of the basic MDR model.

Apparently, the simple MDR model responsible for the pulsar spin-down torque should face a substantial modification \cite{Manchester85,Blandford88,Lyne04,Lyne15, Tong16}. 
One possibility is dedicated to the decoupling of the superfluidity vortex lines in the NS core that transfers the angular momentum into the NS crust; hence, through the variation of the moment of inertia $I$ over time \cite{Allen97}, the braking index can evolve  to depart from the canonical value of 3 \cite{Ho12}; 
another possibility is that the B-field or magnetic angle between the spin and magnetic axis changes with time \cite{Blandford83,Beskin84,Tauris01, Lyne13, Gourgouliatos15,Johnston17,Hamil16};
moreover, when introducing the plasma-filled and non-vacuum magnetosphere, the effects and some possible observed results of pulsar braking are also analyzed~\cite{Melatos97,Contopoulos06, Spitkovsky06, Eksi16}.
Besides the above, the particle wind flows responsible for the pulsar spin-down were also noticed \cite{Goldreich69, Michel69a,Michel69b}, where the electric field accelerates the flow of electrons out of the magnetosphere and takes away the NS angular momentum \cite{Ruderman75, van06}, which can be also taken as
the causes of a pulsar wind nebula (PWN) \cite{Gaensler00, Gaensler06, Hester08}.
In addition, the on--off radio emission phenomena of intermittent pulsar PSR B1931 + 24 \cite{Kramer06}  and  rotating radio transients (RRATs) \cite{McLaughlin06} are interpreted as evidence for the switches of particle wind outflow \cite{van06}.
The braking torques (labeled as T) of these particle flows can be described in the form of $T\propto\Omega$, or the loss rate of kinetic energy of pulsar $\dot{E}\sim L_{f} =  K_{2} \Omega^{2}\propto\Omega^2$, with an undetermined parameter \k2, which has been thoroughly discussed  as a cause of the observed range of the pulsar braking index $1<n<3$  \cite{Harding99,Alvarez04,Xu01, Chen06, Kou15,Lyne15,P19}.
Furthermore, from the aspects of the Crab Nebula, the wind component should be a part of the pulsar braking~\cite{Glendenning96, Hester08}.
According to the particle wind torque model for a pulsar spin-down, the parameter \k2 can be expressed as $K_{2}=\pi\Phi^2/(4c)$, where $\Phi$ is the magnetic flux of the particle flow.
The above parameter \k2 was first studied by Michel \cite{Michel69a}, who extended the relativistic treatment from the solar-wind torque \cite{Weber67}, an expression  that we also apply in our work.
Here, we consider the particle wind flow to modify the MDR model (hereafter referred as MDRW) by introducing the wind flow torque that can independently explain a braking index of $n\sim1$, which can gradually influence the constant braking index of $n=3$, as predicted by the MDR itself.

In this article, the MDRW model is  one of a thought experiment model that can account for the spin-down torque of the Crab pulsar and its analogues.
%we consider both the MDR and particle wind flow (hereafter refered as MDRW) to account for the pulsar spin-down torque by plotting a schematic diagram in Figure \ref{psr}.
Then, we acquire an analytical solution of the spin evolution for the Crab  pulsar in Section~\ref{2}, thereby deriving the evolution formula of the  braking index, characteristic B-field, and characteristic age in Section~\ref{4}.
The {tentative} discussions and conclusions of the evolution evidence for the Crab pulsar are presented in Section~\ref{3}.

\section{Spin Evolution Model Furthermore, Results}\label{2}
In this section, we aim to give an exact solution when the additional wind torque is related to $\Omega^2$, and try to answer the possible evolution of the Crab pulsar.
The reasons for which we chose particle wind outflow are  not only due to the wind component being familiar with the astronomical issue and having been considered in the pulsar braking, as mentioned in the Introduction \cite{Michel69a}, but also because the Crab pulsar wind nebula has been observed, the luminosity of which is comparable to its rotational energy loss rate $\dot E$ \cite{Hester08}. 
Meanwhile, the Gamma ray luminosity by the FERMI-LAT observation is constrained to be approximately 13\% of $\dot E$  for the Crab pulsar \cite{Abdo10}.
Thus, we think that, without taking the particle flow into account, the simple MDR model is incomplete in understanding the spin-down evolution of the Crab pulsar.
%In principle, if the spin-down torque of pulsar is clearly set, the spin evolutionary equation could be solved, including the characteristic B-field, rotational braking index, and characteristic age.
%Borrowing from the historical scheme, such as  the  construction of the analytic solutions of the standard MDR model \cite{Gunn69} that has been written in the pulsar textbooks \cite{Manchester77,Lyne12}, we perform a similar operation procedure.

\subsection*{Analytic Solution of the Pulsar Spin Evolution}
For a pulsar, the loss rate of its kinetic rotational energy is assumed to equate the emission powers contributed by both the MDR ($L_{d}$) and particle wind flow ($L_{f}$), i.e., $ \dot{E} \equiv  L_{d} + L_{f} $, expressed below:
\begin{equation}
-I \Omega \Odot  = K_{1}  \Omega^{4} + K_{2}\Omega^{2},
\label{edot}
\end{equation}
where two undetermined parameters are defined by the MDR model \cite{Gunn69} and particle flow model \cite{Michel69a} for the spin-down of the pulsar, respectively, as mentioned above,
\begin{equation}
K_{1}=2B^2R^6/(3c^3),  {\;\;\;\;} K_{2}=\pi\Phi^2/(4c)\;,
\end{equation}
or, equivalently, Equation~(\ref{edot}) can be simplified as \cite{Lyne15,P19}
\begin{equation}
-\Odot  =  a\Omega^{3} +  b\Omega\;,
\label{odot}
\end{equation}
with the undetermined dipolar parameter $a = K_{1}/I$ and flow  parameter $b = K_{2}/I$, where the condition  $b=0$  or $K_{2}=0$  corresponds to the conventional case of MDR.
As expected, if the two presumed constants $a$ and $b$ were known, then the analytic solution of pulsar rotation might be achieved.
By defining the fraction factors with respect to $\dot{E}$ contributed by the dipolar and flow  components as
\begin{equation}
 {\rm d}=L_{\rm d}/\edot  \;\; ; \;\;  {\rm f}=L_{\rm f}/\edot \;,
\label{df}
\end{equation}
which satisfy the condition of d + f = 1 that is the core assumption of the model.
We proceed by submitting the spin derivative Equation~(\ref{odot}) into the braking index Equation~(\ref{index}), and obtain a relation
\begin{equation}
n = 2{\rm d} + 1 = 3-2{\rm f},
\end{equation}
by which we have  the following expressions
\begin{equation}
{\rm d} = (n-1)/2  \;\; ; \;\;  {\rm f}= (3-n)/2\;.%
\label{dnfn}
\end{equation}

These results demonstrate that the component fraction factors of  pulsar emission are intimately related to the braking index.
For the  Crab pulsar, by employing  the present observed  value (denoted by subscript `o') of the braking index $n_{o}=2.50$,
we obtain $d_{o}=3/4$ and $f_{o}=1/4$, implying that the particle flow as a non-dipolar component  has contributed to 25\% of the total kinetic energy loss.
Furthermore, by Equation~(\ref{df}), we have
\begin{equation}
{\rm d} = a \Omega^3/\dot \Omega  = a\Omega^2\tc \;\; ; \;\;  {\rm f} = b\Omega/\dot \Omega = b\tc \;,%
\label{dftau} \end{equation} where $\tc  = \Omega/\dot{\Omega}$ is defined as
the characteristic age  of pulsar. Necessarily, it is stressed
that we do not use the definition of the characteristic age $\tau =
\Omega/2\dot{\Omega}$ as the magnetic dipole model does
\cite{Lorimer08,Lyne12}, because the braking index of our model
evolves with time and departs  from the canonical value 3
\cite{Manchester85}. After arrangement,  we  obtained the two
parameters of model a and b, respectively, \begin{equation} a ={\rm
d_o}/(\tau_{co}\Omega_o^2) \;\; ; \;\;  b ={\rm f_o}/\tau_{co}\;,
\end{equation} which are calculated by the present spin period and its
derivative of the Crab pulsar as, $a = 2.7\times10^{-16}\,c.g.s$
and $b = 3.2\times10^{-12}\,c.g.s$, corresponding to
$K_{1}=2.7\times10^{29}\,c.g.s$ and
$K_{2}=3.2\times10^{33}\,c.g.s$, respectively.

Similarly, in terms of the fractional  ratio ($\epsilon$) of the particle flow relative to the magnetic dipole (hereafter
referred as the flow--dipole ratio) defined  by Ref.~\cite{Lyne15},   $\epsilon \equiv  {L_{\rm f}}/L_{\rm d}$, we obtain
\begin{equation}
 \epsilon = \frac{\rm f}{\rm d} = \frac{3-n}{n-1}
 = \frac{b}{a\Omega^{2}}\;,
 \label{ep1}
\end{equation}
%simplified  as
\begin{equation}
\epsilon  = (\Omega_m/\Omega)^2 = (P/P_m)^2\;,
\label{ep}
\end{equation}
where $\Omega_m=\sqrt{b/a}=\Omega_o/\sqrt{3}=108.6$ \,rad\,s$^{-1}$ and $P_m={2\pi/\Omega_{m}}=\sqrt{3}P_{o}=57.8\,$ ms, with  the present value of angular velocity  $\Omega_o=188.2$ \,rad\,s$^{-1}$.

\begin{figure}
%\centering
\includegraphics[width=8.6cm]{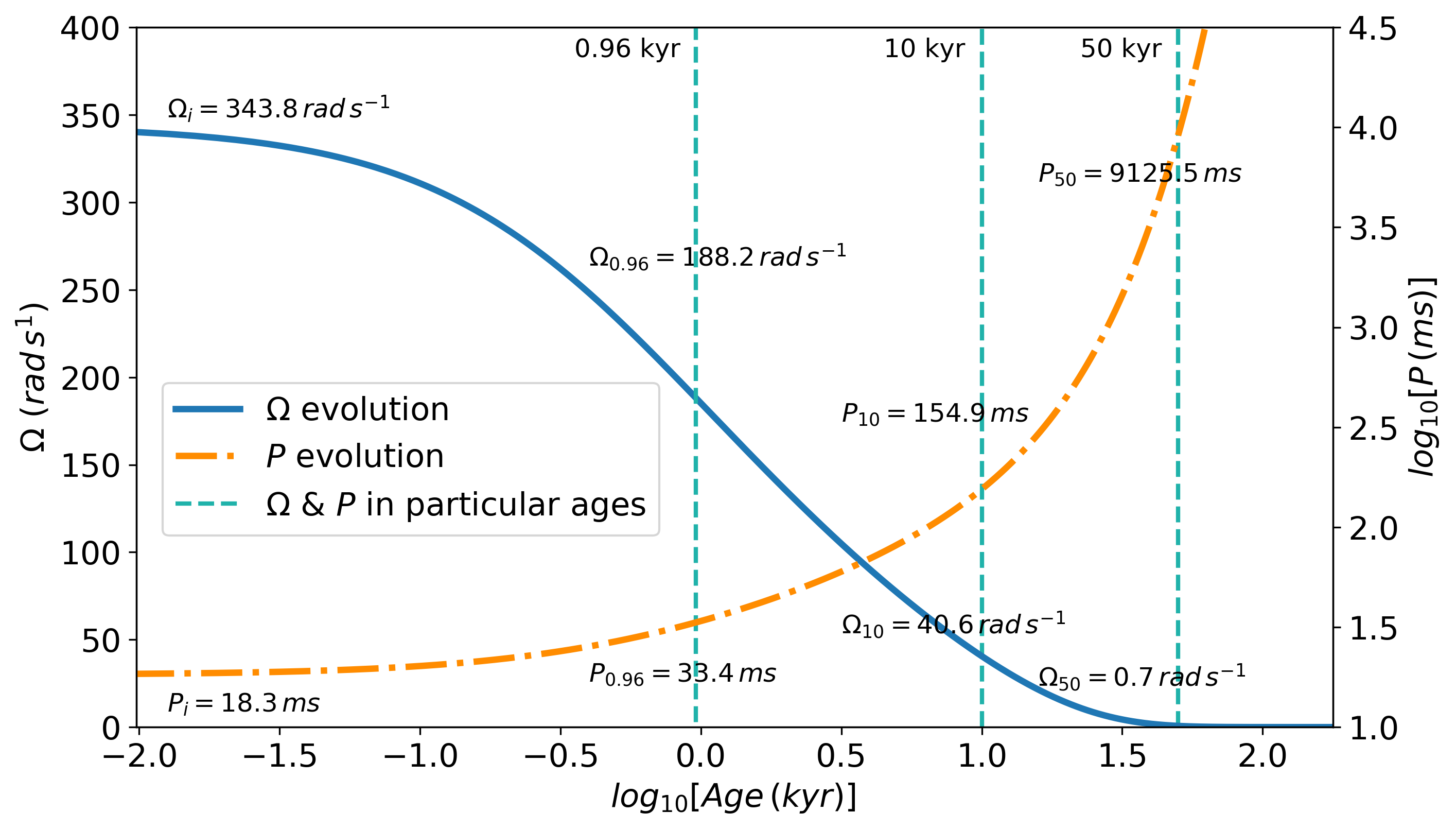}\\
\caption{The evolution of angular velocity ($\Omega$) and spin period ($P$) over real time ($t$).
The solid and dash-dotted lines represent the evolutions of angular velocity and spin period, with the initial values of $\Omega_{i}=343.8$ \;rad s$^{-1}$ and $P_{i}=18.3$ \;ms,  and their particular values at the  real ages of 0.96 kyr, 10~kyr, and 50 kyr are labeled by the three  vertical  dashed lines, respectively, as well as noted in Table~\ref{table1}.
}
\label{omega-p-t}
\end{figure}

Therefore, combining  Equations~(\ref{odot}) and (\ref{ep}),  the differential equation for the spin-down torque evolution can be transformed into the following form
\begin{equation}
\begin{split}
d\epsilon/dt = 2b(1+\epsilon),
\end{split}
\label{epeq}
\end{equation}
and the exact analytic solution of above differential equation can be solved by the integral, then we have % $\epsilon$  expression as
\begin{equation}
\begin{split}
\epsilon  = (1 + \epsilon_{i})e^{2bt} - 1, % = Ce^{2{\rm f_o}t/\tau_o}-1,
\end{split}
\label{epso}
\end{equation}
where $\epsilon_{i}$ is an integral  constant, or the initial value of $\epsilon$ that can be determined by the present spin parameters of the Crab pulsar at $t_o=960$ \,yr (see Table~\ref{table1}, the data from \cite{Lyne15} and ATNF Pulsar Catalogue  \cite{Manchester05}), as  $\epsilon_{i}=\epsilon(t=0)=0.1$.
Thus, the analytic solution of the Crab pulsar's  spin evolution is settled down as below,
\begin{equation}
P=P_{m}\sqrt{\epsilon}=P_{m}\sqrt{1.1 e^{2bt} - 1}\;,
\end{equation}
\begin{equation}
\Omega=\Omega_{m}/\sqrt{\epsilon}=\Omega_{m}/\sqrt{1.1 e^{2bt} - 1}\;,
\end{equation}
which is plotted in Figure \ref{omega-p-t} and shown in Table~\ref{table1}, together with the other related  parameters at  different ages.
To test the validity of this analytic solution Equation~(\ref{epso}), we process it by the Taylor expansion, then  acquire the two special solutions, respectively, for the pure dipolar case  $b \rightarrow 0$ with $\Omega^{-2} = \Omega_{i}^{-2} + 2at $  and the pure non-dipolar case $a \rightarrow 0$ with $\Omega^{2} = \Omega_{i}^{2} e^{-2bt} $, where $\Omega_{i}$ denotes the initial angular frequency. For more details about the derivation of our model, please see Appendix~\ref{appendix d} below.

\begin{table}[H]
\caption{The model-predicted parameters for the Crab pulsar at particular ages. \label{table1}}
%\centering{ {{\bf Table 1.} \; The model predicted parameters for the Crab pulsar at the particular ages}
\begin{adjustwidth}{-\extralength}{0cm}
\begin{tabularx}{\fulllength}{lcccccccccc}
\toprule
{\bf Time (kyr) } &\bf{\boldmath{$\Omega$}(rad\,s$^{-1}$)} &{\bf \boldmath{$-\dot \Omega$} (10$^{-10}$\, rad\, s$^{-2})$} &{\bf \boldmath$P$ (ms)} &{\bf \boldmath $\dot P$(10$^{-13}$\,s\,s$^{-1})$} &{\bf \boldmath$B_{ch}$ (10$^{12}$\, G)} &{\bf \boldmath $\tau_c$ (kyr)} &{\boldmath $n$}& \textbf{f} & \boldmath{$\epsilon$}\\
\midrule
{{0} %MDPI: Please confirm if the bold should be retained.
}      & 343.8  &  $120.4$   & 18.3    & $6.4$   & $3.5$    & 0.9   & 2.82  & 0.09 & 0.10 \\
{{0.96}}   & 188.2  &  $23.9$    & 33.4    & $4.2$   & $3.8$    & 2.5   & 2.50  & 0.25 & 0.33 \\
{{2.98}}   & 108.6  &  $6.9$     & 57.8    & $3.7$   & $4.7$    & 5.0   & 2.00  & 0.50 & 1.00 \\
{{10}}     & 40.6   &  $1.5$     & 154.9   & $5.6$   & $9.6$    & 8.8   & 1.24  & 0.88 & 7.33 \\
{{50}}     & 0.7    &  $0.02$    & 9126    & $290$   & $334$    & 9.967   & 0.9999   & 0.9999 & 10,000\\
\bottomrule
\end{tabularx}
\end{adjustwidth}

\footnotesize{{\bf Note: } The Crab pulsar real  B-field $B = 3.3\times 10^{12}$ G and maximum characteristic age $\tau_{cmax}=\tau_{co}/f_{o}=b^{-1}\simeq10$~kyr.}

\end{table}

\vspace{-12pt}

\section{Magnetic Field and Braking Index} \label{4}

The pulsar observable quantities such as the characteristic B-field, rotational braking index, and characteristic age are all determined by the spin periods and their derivatives.
Therefore, their evolutions can be derived in terms of the spin period solution, which are described in the following.

\subsection{Growth of Characteristic Magnetic Field}
\begin{figure}[H]
%\centering
\includegraphics[width=8.6cm]{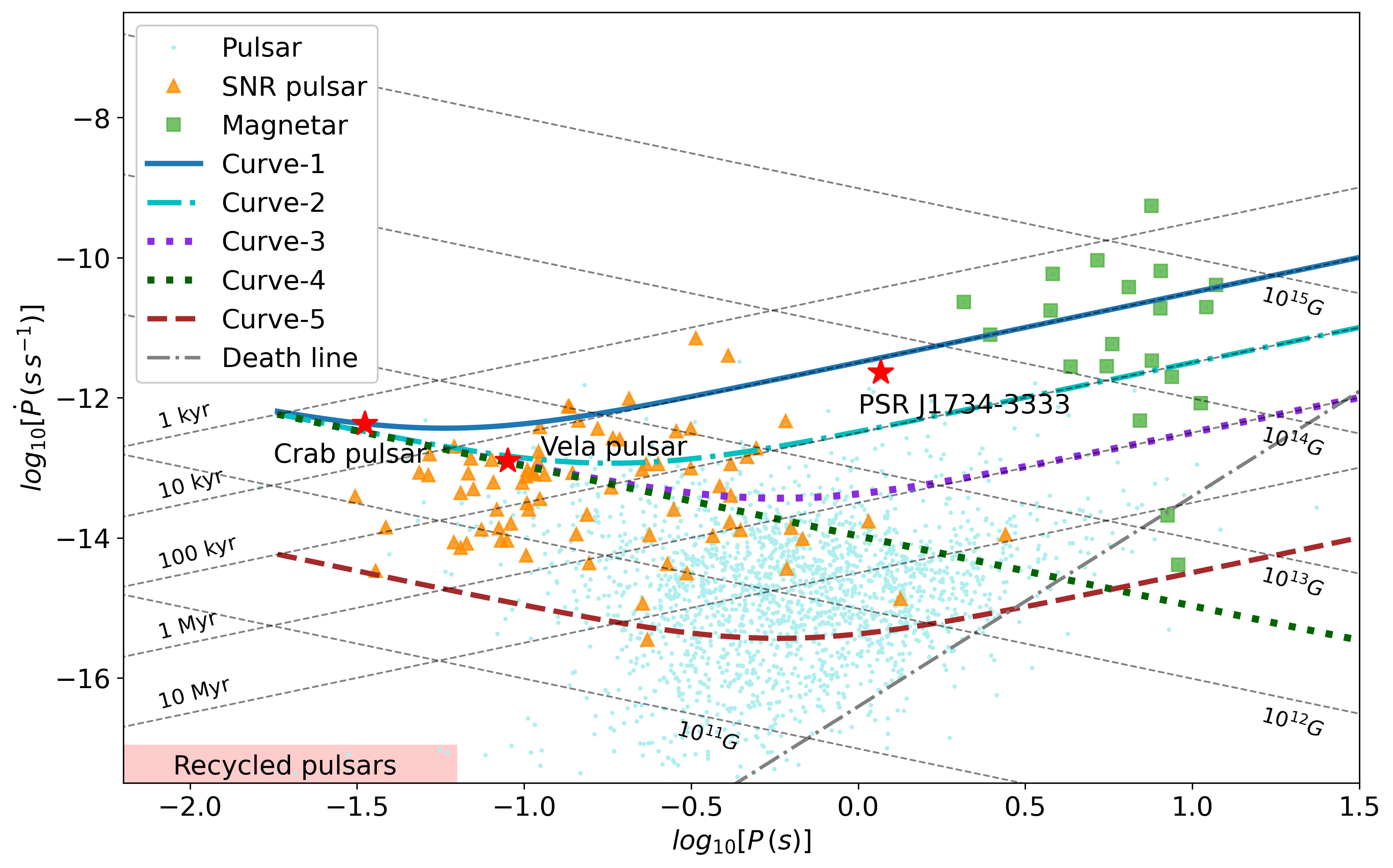}\\
\caption{
Diagram of the derivative of the spin period versus spin
period ($P-\dot P$ diagram). The solid curve (Curve-1) stands for
the $P-\dot P$ evolutionary track for the Crab pulsar with
$B_1=B=3.3 \times 10^{12}$ G and $b_1=b=3.185\times10^{-12}$ 
s$^{-1}$.
The dash-dotted line (Curve-2) and dotted line (Curve-3)
represent the situations with the conditions $B_2=B$ and
$b_2=0.1b$ and $B_3=B$ and $b_3=0.01b$, respectively.
The dark-green-dotted line (Curve-4) is the MDR evolution curve with the parameters $B_4=B$ and $b_4=0$.
The dashed line (Curve-5) is plotted with the values of $B_4=0.1B$ and
$b_5=0.01b$. The light-dash-dotted line represents the death line
for the pulsar ceasing its radio emission
\cite{Ruderman75,Bhattacharya91}. The light-dashed
lines stand for the  different characteristic age and B-field,
respectively. The data   were taken
from the ATNF Pulsar Catalogue  \cite{Manchester05}.
The red stars represent  particular pulsars,
including the Crab pulsar, the Vela pulsar, and PSR J1734-3333.
Below the symbol ``Recycled pulsars'' are those experienced the binary accretions \cite{Bhattacharya91, Phinney94,Zhang06}.
}
\label{p-pdot}
\end{figure}

When the particle flow recedes to null  ($ \epsilon=0$, $ \rm f=0$ and $\rm d=1$), the MDRW model returns to the conventional dipolar case.
If the particle flow  switches on, then the loss rate of kinetic energy $\edot = L_{\rm d} (1 + \epsilon) $  can derive the characteristic B-field  as  $B_{ch} \propto \sqrt{P\dot{P}}$, but the modification factor of $(1 + \epsilon) $ to the dipolar case is introduced.
Thus, the relation between the characteristic B-field  ($B_{ch}$) and the real B-field strength ($B$) is given by
\begin{equation}
 B_{ch} =  B \sqrt{1+\epsilon} \;,
 \label{bch}
\end{equation}
which infers  that $B_{ch}$ is not  a constant  and  increases  with the spin evolution.
In other words, the real B-field of the Crab pulsar remains a constant and should be calculated by both the characteristic B-field and the flow--dipole ratio  factor $\epsilon$,  expressed in terms of the pulsar parameters at the present time
\begin{equation}
B = B_{cho}/\sqrt{1+\epsilon_{o}} = 3.3\times 10^{12}  G\;.
 %\label{bch}
\end{equation}
As shown in the diagrams of $P-\dot P$ (Figure \ref{p-pdot}), the $B_{ch}$ of the Crab pulsar increases by two orders of magnitudes to $\sim10^{14}$ \,G when the real time goes to $\sim50$ \,kyr.
Thus, if our model is correct, then this result may indicate that the Crab pulsar may become a high characteristic B-field pulsar in the future, which could answer the previously proposed conjectures~\cite{Lyne04,Espinoza11} that some pulsars may move to the high B-field range with evolution.
In detail, the $P-\dot P$ evolutionary track of the Crab pulsar shows that it moves from its birth place with the initial spin period of $ P_{i} = 18.3$ \,ms, derivative of $\dot P_{i} = 6.4\times 10^{-13}$ \,{s\,s$^{-1}$} %MDPI: Please confirm this unit.
, and the initial characteristic B-field of $B_{chi} = 3.5\times 10^{12}$ \,G.
Then, the path passes through the locations of the Vela pulsar (PSR B0833-45, $n=1.4$, $P=89$ \,ms, $B_{ch}=3.38\times10^{12}$ \,G) \cite{Lyne15} and the high B-field radio pulsar PSR J1734-3333 ($n=0.9\pm 0.2$,  $P=1.17$ \,s, $B_{ch}=5.2\times10^{13}$ \,G), arriving at those of the ``magnetar'' population, if there were no other factors influencing the pulsar spin down.
It is noted that the real B-field of the Crab pulsar is still maintaining its same initial value of $B=3.3\times10^{12}$ \,G,
{ and the characteristic field might appear much higher than the actual one if interpreted with the consideration of a smaller braking index.}
Here, it is clarified that we do not suspect the potential super-strong B-fields of magnetars, since the emission properties of both the Crab pulsar and most magnetars are very different; therefore, our simple but thoughtful  model could not be automatically applied to those special sources of soft gamma-ray repeaters (SGRs) and anomalous X-ray pulsars (AXPs) \cite{Duncan92,Thompson93, Ferrario08, Kaspi17, Esposito21}  with extremely intense high-energy outbursts (more explanations  are noted  in Appendix~\ref{appendix e}).

In Figure \ref{p-pdot}, the tendency of the  $B_{ch}$ curve evolution can be understood by its evolutionary equation whereby the slop parameter k of the $B_{ch}$ curve is almost null at the early age, horizontally going along the constant B-field line (\emph{n} = 3), and slop k gradually approaches unity while  the $B_{ch}$ curve moves into the ``magnetar'' population (\emph{n} = 1). 
Equivalently, in the $P-\dot{P}$ diagram, the slop of the $B_{ch}$  curve  moves from k = $-$1 to 1, which is similar to the predicted route from the Crab pulsar via PSR J1734-3333 to the high B-field population \cite{Espinoza11}, corresponding to the cases of the braking index of \emph{n} = 3 and \emph{n} = 1, respectively, \cite{Johnston17}.
For a further illustration, the evolution of the characteristic B-field $B_{ch}$ with the low values of the B-field and flow parameter b is also plotted, and its evolutionary path covers the vast population of normal pulsars in the $P-\dot P$  diagram.
%Meanwhile, as a reference, we also notice the samples of the measured B-fields by the cyclotron absorption lines of X-rays for the accreting NSs in HMXBs, which centers at a narrow regime around $B = 3\times 10^{12}$ G \cite{Truemper78,Ye19}.}

%\begin{figure}
%\centering
%\includegraphics[width=7.8cm]{B-P.png}\\
%\caption{Diagram of the B-field versus spin period ($P-B$ diagram), with the same label meanings as those of Figure 2.}
%\end{figure}

\subsection{Decay of Braking Index }
\begin{figure}[H]
%\centering
\includegraphics[width=8.6cm]{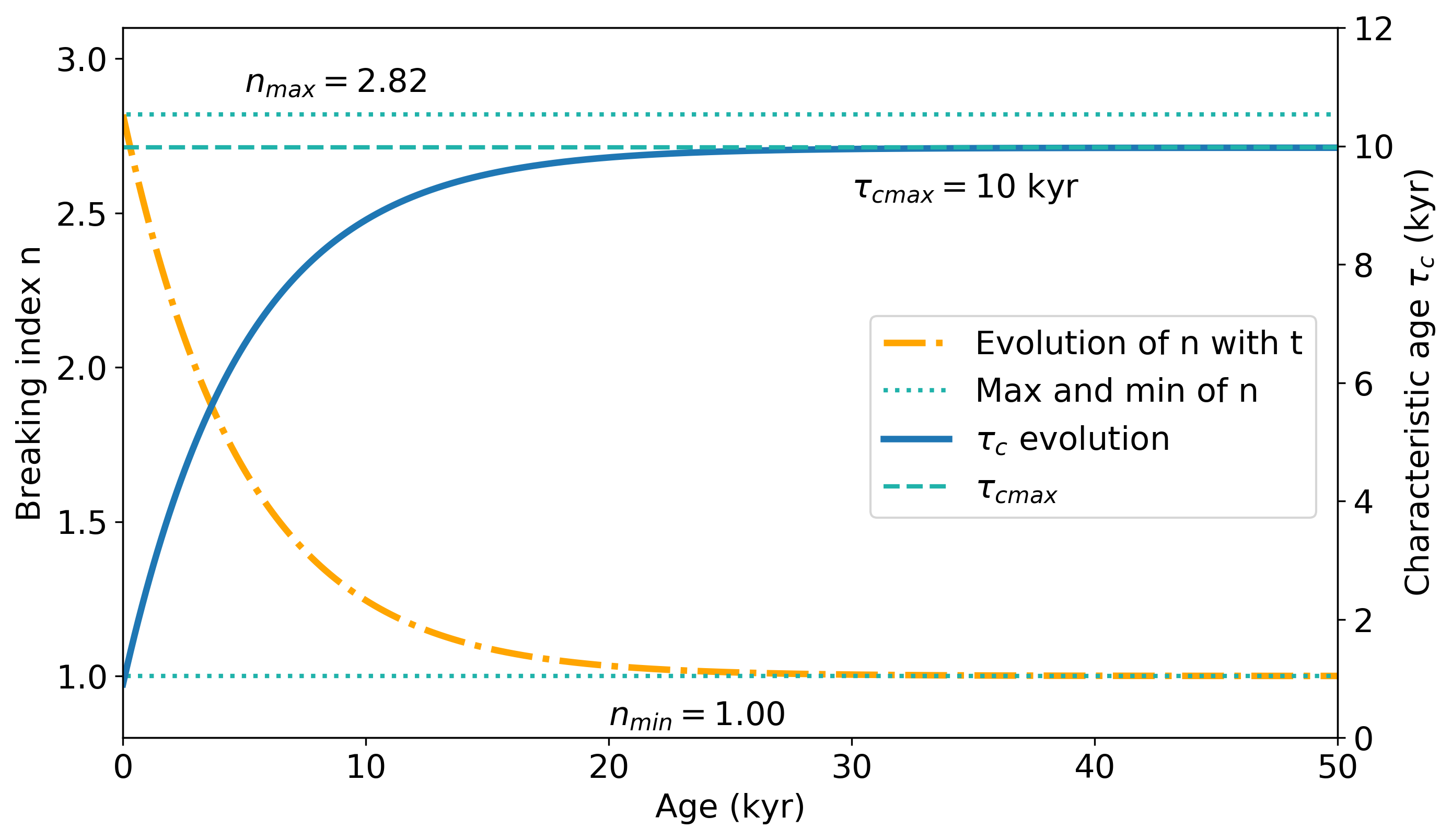} \\
\caption{Evolution of braking index ($n$) and characteristic age ($\tau_c$) over time ($t$) for the Crab pulsar.
The chain-dotted and solid curves stand for the  evolution of the braking index and characteristic age, respectively, while the dotted lines indicate the  upper/lower limit of a braking index \mbox{$n_{max}$ = 2.82}/$n_{min} = 1$ and  the dashed line and dotted line denote the upper/lower  limit of characteristic age of  10/0.9~kyr, respectively.}
\label{n-tau-t}
\end{figure}

Currently, approximately eight radio pulsars are measured by the stable braking index with long-term observation \cite{Lyne15,Espinoza17}, and most of them are randomly and quite evenly spread in the expected range between 1 and 3.
These phenomena are consistent with the predictions of the MDRW model  \cite{Michel69a}.
Therefore, we can obtain the evolutionary equations of the braking index by solving Equation~(\ref{ep1}), that is
\begin{equation}
n  =  3 - \frac{2\epsilon}{1+\epsilon}.
\label{n}
\end{equation}
With evolution, the flow--dipole ratio $\epsilon$ ranges from 0 to $\infty$, which is corresponding to the flow-total fraction factor from 0 to 1, implying an index  ranging from  3 to  1 owing  to the two extreme cases of dipolar or non-dipolar domination, which are plotted in Figure \ref{n-tau-t}, where we notice the decay of the braking index with time or spin-down.
Notably, the braking index of the Crab pulsar decreases from 2.51 to 2.50 over 45 years until the present stage, which is consistent with the observational results \cite{Lyne15}.

Meanwhile, some other evidence may also be consistent with the above relation, such as the specific pulsars mentioned in Figure \ref{p-pdot}, such as the Vela pulsar ($n=1.4$) and PSR J1734-3333 ($n=0.9\pm0.2$), which are seemingly exhibited  as the later evolutionary phase of  the Crab pulsar \cite{Lyne15}.
If we assume that our thought experiment MDRW model has merit, then it can easily explain  values such as these.
Moreover, for the index $n=2$, corresponding to $\epsilon=1$ as shown in Table~\ref{table1}, it represents a transitional point at which  the dipolar and non-dipolar  balance appears, with f=d=0.5 and $ P=P_{m}$, $\Omega=\Omega_m$  as listed in Table~\ref{table1}, implying that the two components account for the same proportion of total radiation.

\subsection{Evolution of Characteristic Age }
The characteristic age is usually used to estimate the approximate age of young, regular middle-aged, or old recycled pulsars, although it is known that this may not be accurate enough \cite{Camilo94, Lyne96, Lorimer08, Cui21}.
It is interesting to examine the evolution of $\tau_c$  in our MDRW model, as exhibited below:
\begin{equation}
\tau_c  %  \frac{f}{b}
= \frac{\epsilon}{b(1+\epsilon)} = \frac{4\tau_o\epsilon}{1+\epsilon}\;,
\end{equation}
the evolution of which is shown in Figure \ref{n-tau-t}.
Strikingly, for the Crab pulsar,  $\tau_c$ has an upper limit  of $\tau_{cmax}=1/b=4\tau_o\simeq10$ \,kyr in our model, which corresponds to the flow-total fraction f = 1 or flow--dipole ratio  $\epsilon  \to \infty$, meaning that the characteristic age of a pulsar will not increase forever with  evolution.
In addition, the minimum value of the characteristic age  $\tau_{cmin}= 0.9$ \,kyr is obtained by the initial condition.
That is to say, $\tau_c$  varies  within a limited range of (0.9--10) kyr,
therefore,  it is a  coincident event  for the Crab pulsar  that the characteristic age is close to the real age at the present stage.
This result reminds us that the characteristic age cannot represent the real ages of pulsars  in general, if the particle flows share a big portion in the total radiation.
For a clear view, $\tau_c$ values at various time stages  are also  shown in Table~\ref{table1}.

\section{Discussion and Conclusion}\label{3}

\subsection{Measurement of Braking Index}
For pulsar braking index measurement, stochastic timing variations by the NS spin instabilities exist which may be related to the spin-down torque, so it is not easy to acquire the precise braking index \cite{Chukwude10,zhang12}, 
and the elimination of noise is very important~\cite{Cordes80}.
To date, only eight pulsars have reliable braking index values, because it is hard to obtain an accurate second derivative of the period.
Generally speaking, there are three difficulties~\cite{Archibald16}
which need long-term and continuous monitoring by radio telescopes: namely the Crab pulsar, which  has a \mbox{45-year} observation \cite{Lyne15};
the pulsar timing irregularities including glitches, which may significantly affect the measurement accuracy,
and the effect of the random walk related to the micro-glitches which may be supposed to be the main source for the impossible accurate braking index measurement;
and some other aspects, which may have less potential influence but still need to be mentioned, including interstellar scatting, scintillation, etc.
%even the effect of the pulsar proper motion is also required to be removed \cite{Bisnovatyi-Kogan93}.}
Wavelet analysis is a powerful time--frequency analysis method, like Fourier transform, which is particularly suitable for processing unstable signals such as complex noise or sudden changes in the data.
Although this method is effective and has been applied in some fields of astronomy  such as white dwarf \cite{Ghaderpour20} and pulsar \cite{Jiang17,Zhong07}, it has not yet been widely used in pulsar data processing from our knowledge.
Perhaps the method of wavelet analysis will be helpful for the measurement of a braking index in the future \cite{Jiang17}.
Furthermore, due to the James Webb Space Telescope having been successfully launched and operated \cite{Gardner06}, more details of the Crab nebula will be discovered and analyzed.
Since the magnetic field of NS has been estimated by its nebula \cite{Boccaletti65, Sabbata65} before the first pulsar was discovered, the surrounding environment is quite important for studying the characters of its central engine.
Particularly for the pulsar wind, accurate observations and studies will enhance our knowledge of the torque of particle flow and additionally constrain and test our MDRW model.

\subsection{Limitations and Assumptions of MDRW Model}

Firstly, our MDRW model about pulsar spin-down has two parts, namely vacuum MDR component and wind flow.
Vacuum around the pulsar should be a simple case (the modification may be due in the future, and more explanations are listed in Appendix~\ref{appendix e}), and the wind flow is borrowed from a previous study \cite{Michel69a}.
Then, we assume constant a and b parameters, which mean that there is no change in the real B-field and magnetic inclination angle in our model.
Therefore, variations in  the above parameters will alter our conclusions, as shown in Figure \ref{p-pdot} (various initial B-field and parameter b).
Meanwhile, if the multipole B-field exists \cite{Petri15,P19}, the results of the paper should also be changed.
After that, the Crab pulsar perhaps cannot represent all pulsars, because the braking index of PSR J1640–4631 is higher than 3 ($n=3.15\pm0.03$) \cite{Archibald16}, meaning that the other mechanism should be possible, and our model is probably not suitable for the apparent non-Crab-like group.
While for most pulsars with accurate measurement \cite{Lyne15}, the braking index is less than 3, such as PSR B0540-69 ($n=2.14\pm0.009$) \cite{Livingstone07}, the Vela pulsar PSR B0833-45 ($n=1.4\pm0.2$) \cite{Lyne96}, PSR J1119-6127 ($n=2.684\pm0.002$) \cite{Weltevrede11}, PSR B1509-58 ($n=2.839\pm0.001$) \cite{Livingstone07}, PSR J1734-3333 ($n=0.9\pm0.2$) \cite{Espinoza11}, PSR J1833-1034 ($n=1.8569\pm0.001$) \cite{Roy12}, and PSR J1846-0258 ($n=2.65\pm0.1$) \cite{Livingstone07}, our model may be suitable for these sources.
Besides these, some other pulsars also have potential measurements \cite{P19} such as PSR J1208-6238 ($n=2.598$) \cite{Clark16}.
%and PSR J1803-2137 ($n=1.9$), and PSR J1826-1334 ($n=2.2$)
Thus, our model is limited to Crab-like pulsar with strong wind, while the population of pulsars can be diverse \cite{Cui21}.
Further questions and problems about MDR and MDRW are addressed in Appendices A-E as listed below.

\subsection{Theoretical Interpretation and Observation Evidence }
Based on both the MDR and particle wind flow contributions to the NS spin-down torque, this study has obtained an exact analytic solution for a pulsar spin period evolution, thereby  applied to the Crab pulsar to give rise to three primary outcomes, i.e.,
(1) the substantial enhancement of the characteristic B-field from $\sim$3.3 $\times$ 10$^{12}$ G to  $\sim$10$^{14}$ G in 50~kyr, {while the real B-field remains unchanged;} 
(2) the decay of the braking index from 2.51 to 2.50 over \mbox{45 years} to the present day and from 2.82 to 1 for a long-term evolution;
and (3) the saturation of a characteristic age at 10 kyr with the continuation of time.%English editor: Please ensure that the meaning has been retained.

These {tentative} conclusions are consistent with the observations, parts of which were also noticed by \mbox{astronomers~\cite{Lyne04,Lyne15}}.
Although the characteristic B-field increases approximately hundred times, the real B-field does not change at all, and hence, the Crab pulsar could possibly evolve to a high characteristic B-field pulsar such as PSR J1734-3333 \cite{Espinoza11,Espinoza22}.
Therefore, we do not think that the older Crab pulsar could evolve to a real magnetar with a super-strong true B-field \cite{Duncan92}, and our model is unsuitable for the magnetars due to their violent high energy emission outbursts (see Appendix~\ref{appendix e}).
%On the reason why astronomers have not discovered radio pulsars with the higher B-field than $10^{14}\,G$  may be that: such a high characteristic B-field pulsar still has a real B-field of $\sim 10^{12}\,G$, and a long spin period of $\sim$  10 s makes it locate beneath  the death-line of pulsar, where the radio emission ceases its function \cite{Ruderman75}.
%Perhaps,
% the particle flow efficiency decays in several 10 kyrs, making
%the characteristic age of pulsar longer than 10 kyrs, while the
%high $B_{ch}$  decreases to $\sim 10^{13}$ Gauss, entering into
%the normal pulsar population. In Figure 2, the influence of
%the low value of b-parameter on the pulsar evolution track is
%plotted.
%
%Meanwhile, we predict the existence of  the radio pulsar with the
%characteristic B-field higher than $\sim 10^{14}$ Gauss, then this
%peculiar object could be rare case, which needs a more powerful
%telescope like 500-m aperture spherical radio telescope (FAST)
%\cite{Li18} to observe its fainter emission in the future.
Now, a question arises: do all pulsars evolve to high characteristic B-field pulsars like the Crab pulsar potentially does?
Our answer is  ``No''.
If we consider the lower value of particle flow contribution (e.g., the low value of $K_{2}$ or b)
than that of the Crab pulsar (see Curve-5 in Figure \ref{p-pdot}), we find that the $P-\dot{P}$ evolutionary path goes to the region of normal pulsars ($\sim$10$^{12-13}$ \,G).
In other words, most radio pulsars seem to follow almost constant B-field routes over millions of years.
Thus, the different particle flow contributions may account for the distribution of pulsars in the $P-\dot{P}$ diagram, which reminds us to relate them to the two types of pulsars with and without the SNRs \cite{Cui21}.

%Besides the enhancement of the characteristic B-field, the characteristic age  and real age of  pulsar  are different by the DC model.
%As known, from  the MDR model, the characteristic age is bigger than the real age, which has risen lots of debates and discussions \cite{Camilo94,Lorimer08}.
%Several observations showed that the characteristic age of pulsar is
%far from the expansion age of its supernova remnant or proper motion age \cite{Espinoza17, Espinoza22, Johnston21}, which are usually taken as the indicators of the real ages.
%In addition, it was noticed that the increasement of the characteristic age and real age may be not synchronized \cite{Lyne04}.
%However, our model pointed out that the discrepancy between two  ages represents the significant  contribution of  the particle flow component.
%For the Crab pulsar, at the time scale of several thousands of years, there is no big difference  between the two ages  by our model, however a much more deviation occurs after 50 kyrs while the characteristic age reaches a maximum value of 10 kyrs and keeps unchanged afterwards.
%Hence, we conclude that the characteristic age is not a good indicator for the real age if the sufficient particle wind flow exists, which is about 25\% of the rotational kinetic energy loss rate for a Crab pulsar.
%thus the MDR model should be incomplete for describing the evolution of  the Crab-like pulsars.

As known from the MDR model, the characteristic age is bigger than the real age, from which many debates and discussions have arisen \cite{Camilo94,Lorimer08}.
Several observations have  shown that the characteristic age of pulsar is far from the expansion age of its supernova remnant or proper motion age \cite{Espinoza17, Espinoza22, Johnston21}, which are usually taken as the indicators to estimate their real ages.
In addition, it was noticed that the increment in the characteristic age and real age may be not synchronized \cite{Lyne04}.
However, our model pointed out that the discrepancy between the two ages may be due to the significant contribution of the particle flow component.
Hence, we concluded that the characteristic age of pulsar is not a good indicator of the real age if sufficient particle wind flow exists.
The wind component is approximately 25\% of the rotational kinetic energy loss rate for the Crab pulsar ($\dot{E}=5 \times 10^{38}$\, erg/s) inferred by the braking index \emph{n} = 2.5 \citep{Lyne15}.
In addition, the observed luminosity of the Crab nebula is approximately  $1.3\times10^{38}$ \,erg/s, which is also noticed to be 26\% of the total luminosity of the Crab pulsar \citep{Hester08}.
Thus, the values of ratio (f) between the wind  flow and total energy loss rate are consistent based on the two observations, the pulsar timing and nebula luminosity, which may favor  our model to some extent.

Furthermore, we predicted the imaginable future parameters of the Crab pulsar based on the MDRW, while the torque of wind dominates its braking, e.g., after 20 kyr, its braking index is about 1, the spin period will slow down to $\sim$$0.6$ \,s, and the characteristic B-field will reach $\sim$$3.5\times10^{13}$ \,G.
Then, another question arises, namely  that of whether there a source to proof this evolution path?
The answer is a strong ``maybe'', considering the assumptions of our model.
Currently, only a few radio pulsars have the reliable braking index measurements as mentioned above, among which only one pulsar has n very close to 1, namely PSR J1734-3333, holding with $n \sim 1$,  $P = 1.17 \,$ s, and $B_{ch}\sim 5.2\times 10^{13}$ \,G \citep{Espinoza11}, which are close to the future features of the Crab pulsar.
For the pulsar PSR J1734-3333, both the estimated SNR age \mbox{(23 kyr) \citep{Ho12}} and proper motion age (45--100 kyr) \citep{Espinoza22} are much older than its characteristic age (8.1 kyr), which might hint at the fact that it may be saturated like the Crab pulsar probably is.
If we calculate the ratio parameter $\epsilon$ between the wind flow and MDR with the age of 23~kyr and initial magnetic field of $3.3\times10^{12}$ \,G by Equation~(\ref{bch}), we obtain   $\epsilon \sim 248$, implying that the torque of the wind flow is higher than that of its MDR for  PSR J1734-3333.
Meanwhile, the corresponding braking index of 1.008 is obtained by Equation~(\ref{n}), which satisfies its observed constraint of $n=0.9\pm0.2$.
Moreover, the X-ray luminosity of this pulsar is weak~\citep{Espinoza11}, which may be due to the fact that the high energy particles do not emit in our line of sight, or PSR J1734-3333 has insufficient overall energy (since its \mbox{$\dot E \sim 5.6\times10^{34}$ \,erg/s} is four orders of magnitude lower than the Crab pulsar).
Thus, PSR J1734-3333 is a possible candidate for its contribution of wind flow in braking torque as it is significantly higher than that of MDR.

%%%%%%%%%%%%%%%%%%%%%%%%%%%%%%%%%%%%%%%%%%
\vspace{6pt}

%%%%%%%%%%%%%%%%%%%%%%%%%%%%%%%%%%%%%%%%%%
%% optional
%\supplementary{The following supporting information can be downloaded at:  \linksupplementary{s1}, Figure S1: title; Table S1: title; Video S1: title.}


%%%%%%%%%%%%%%%%%%%%%%%%%%%%%%%%%%%%%%%%%%
\authorcontributions{C.-M.Z. and X.-H.C.  completed the model derivation, wrote the paper, and outlined   this study.
C.-M.Z., D.L., D.-H.W., N.W., S.-Q.W.,  B.P., and W.-W.Z. conceived and refined the physical interpretation of the results.
X.-H.C.  and J.-W.Z. are responsible for writing programs and analyzing data.
X.-H.C.,   Y.-Y.P., and  Y.-Y.Y. participated in drawing figures and table.
All authors have read and agreed to the published version of the manuscript.}

\funding{This work is supported by NSFC (Grant No. 11988101, No. U1938117,  No. U1731238, No.
12163001, No. 11725313, and  12130342), the International
Partnership Program of CAS grant No.
114A11KYSB20160008, the National Key R\&D Program of China No.
2016YFA0400702, and the Guizhou Provincial Science and Technology
Foundation (Grant No. [2020]1Y019).}

\institutionalreview {{ "Not applicable" for studies not involving humans or animals. }}%{In this section, you should add the Institutional Review Board Statement and approval number, if relevant to your study. You might choose to exclude this statement if the study did not require ethical approval. Please note that the Editorial Office might ask you for further information. Please add “The study was conducted in accordance with the Declaration of Helsinki, and approved by the Institutional Review Board (or Ethics Committee) of NAME OF INSTITUTE (protocol code XXX and date of approval).” for studies involving humans. OR “The animal study protocol was approved by the Institutional Review Board (or Ethics Committee) of NAME OF INSTITUTE (protocol code XXX and date of approval).” for studies involving animals. OR “Ethical review and approval were waived for this study due to REASON (please provide a detailed justification).” OR “Not applicable” for studies not involving humans or animals.} 

\informedconsent {{"Not applicable" for studies not involving humans. }}%{Any research article describing a study involving humans should contain this statement. Please add ``Informed consent was obtained from all subjects involved in the study.'' OR ``Patient consent was waived due to REASON (please provide a detailed justification).'' OR ``Not applicable'' for studies not involving humans. You might also choose to exclude this statement if the study did not involve humans. Written informed consent for publication must be obtained from participating patients who can be identified (including by the patients themselves). Please state ``Written informed consent has been obtained from the patient(s) to publish this paper'' if applicable.}

\dataavailability{The data underlying this article are taken from ATNF Pulsar Catalogue, available at \url{https://www.atnf.csiro.au/research/pulsar/psrcat/} {(accessed on 1 March 2022)} %MDPI: Please add the access date (Format: Date Month Year). e.g., (accessed on 1 January 2020).
.
The description of the total number of pulsars is based on the following websites: \url{https://www.atnf.csiro.au/research/pulsar/psrcat/} {(accessed on 1 March 2022)}, \url{http://zmtt.bao.ac.cn/GPPS/} {(accessed on 1 March 2022)}, and \url{http://groups.bao.ac.cn/ism/CRAFTS/} {(accessed on 1 March 2022)}.}

\acknowledgments{We are grateful to Richard Strom for carefully reading the manuscript. 
Meanwhile, we sincerely thank all referees for the meaningful comments and suggestions, which have significantly improved the quality of this paper.}

\conflictsofinterest{The authors declare no conflict of interest.}

%%%%%%%%%%%%%%%%%%%%%%%%%%%%%%%%%%%%%%%%%%
%% Optional
%\sampleavailability{Samples of the compounds ... are available from the authors.}

%% Only for journal Encyclopedia
%\entrylink{The Link to this entry published on the encyclopedia platform.}

%\abbreviations{Abbreviations}{The following abbreviations are used in this manuscript:\\

\appendixtitles{yes} % Leave argument "no" if all appendix headings stay EMPTY (then no dot is printed after "Appendix A"). If the appendix sections contain a heading then change the argument to "yes".
\appendixstart
\appendix
\section[\appendixname~\thesection]{Basic Information for MDR and Wind Component} \label{appendix a}
%\subsection[\appendixname~\thesubsection]{}

%\section*{}

 As one of the simplest  theories to describe pulsar emission, the MDR model was developed by Gunn and Ostriker \cite{Gunn69, Ostriker69} based on  pioneer NS proposals \cite{Pacini68,Gold68} soon after the discovery of the first pulsar \cite{Hewish68}, which was taken as a ``standard version” for the evolution and emission mechanism in the pulsar and NS text books \cite{Manchester77,Shapiro83, Meszaros92, Camenzind07, Ghosh07, Becker09, Lyne12, Lorimer12}.
It has been a successful model in pulsar astronomy for over 50 years, and is usually applied on the pulsars powered by their rotational energy loss.
Moreover, the characteristic magnetic field of pulsar is estimated by the MDR, which has been popularly used by astronomers and astrophysicists worldwide and in the data base of the ATNF Pulsar Catalogue \cite{Manchester05}.
In the MDR model, the pulsar is regarded as a rapidly rotating rigid body with a strong magnetic field located in a vacuum environment, so the corresponding physical picture of MDR can be regarded as a simple magnetic dipole form.
From the mathematical expression, the momentum is proportional to the angular velocity $\Omega^4$, and can be further expressed as $I\Omega \dot \Omega$.
Therefore, if we want to know how the pulsar evolves with time under MDR, equating the differential equation related to $\Omega$ and $\dot \Omega$ can be obtained as made by the ``standard MDR model''.

However, as illustrated in this work, the simple MDR cannot explain some phenomena, such as the  braking index lower than 3 (\emph{n} = 3 is expected by MDR).
Thus, one of modified methods is introduced by the contribution of particle flow during the pulsar \mbox{spin-down~\cite{Manchester85,Blandford88}}.
After introducing the flow, the pulsar braking is not only caused by MDR but also by the angular momentum loss due to particle wind.
Mathematically, the latter effect can be considered to be  proportional to $\Omega^2$ \cite{Alvarez04, Lyne15}.
Therefore, when considering the evolution under the MDRW model, we can also solve the differential equation between $\Omega$ and $\dot \Omega$ (Equation (\ref{edot}) in the manuscript), and fortunately, acquire an exact analytic solution.

%\section*{Appendix B: }
\section[\appendixname~\thesection]{Characteristic B-Field, Braking Index, and Characteristic Age} \label{appendix b}
These are important parameters for pulsar evolution, and there are some explanations for these three parameters.
To begin with, the characteristic B-field is calculated based on the MDR model, which is derived by the period and period derivative.
However, this value may not be the real B-field of the pulsar, and only some pulsars in the high-mass X-ray binaries have their measured value inferred by the cyclotron absorption \cite{Ye19}.
Next, it is known that the rotation for normal pulsars are slowing down, so the braking index accurately describes the spin-down of the pulsar \cite{Johnston99}.
Under the MDR model, the braking index is obtained to be 3, while from the aforementioned observations, we know that none of the measured n is 3, and most of them are in the range of 1--3.
Then, due to the braking of the pulsar, the characteristic age represents the time for the pulsar to evolve to its present spin period with a consistent rate. Therefore, astronomers used to take the ratio between the period and period derivative to estimate the approximate age, but now this needs to change.
For example, the Crab pulsar, the real age of which is $\sim$960 yr, and whose characteristic age is $\sim$2500 yr (under MDR, this value changes to 1260 yr).
The real age and characteristic age are not consistent.

%\section*{Appendix C: }
\section[\appendixname~\thesection]{Parameter List for the Crab Pulsar and MDRW Model} \label{appendix c}

We added a parameter table for a better understanding on MDRW model that we used as follows.
The coefficients in the model have already been mentioned in Introduction section, such as the moment of inertia ($\sim$10$^{45}$ gcm$^2$) and radius ($\sim$$10$ km) of the NS we applied.
These coefficients are widely used in pulsar astronomy as well as in the ATNF Catalogue and Pulsar textbook. Based on these, the mass of NS can be estimated by $I\approx 2/5MR^2$, so the mass is approximate $\sim 1.3M_{\odot}$ ($M_{\odot}\sim2\times10^{33}$ g).
However, this result is under the assumption of a uniform sphere.
If considering an inhomogeneous sphere, a modification coefficient of less than 1 needs to be multiplied (e.g., 0.7 or 0.8), so the calculated mass is $\sim 1M_{\odot}$.
The above values are consistent in our derivation and analysis, and even though they slightly change, they will not affect our results.

\begin{table}[H]
%\centering{ {{\bf Table 2.} \; Parameter list for our model and the Crab pulsar.}
\caption{Parameter list for our model and the Crab pulsar $^a$.}
\newcolumntype{C}{>{\centering\arraybackslash}X}
\begin{tabularx}{\textwidth}{CCC}
\toprule
\textbf{Parameter}	& \textbf{Value}	& \textbf{Ref.}\\
\midrule
$I$ \,(g\,cm$^2$)   & $\sim$$10^{45}$      &  \cite{Haensel07, Glendenning96, Lattimer04, Lorimer12} \\
$R$ \,(km)   &$\sim$10       & \cite{Becker95, Lorimer12}\\
$M\,(M_{\odot})$   &$\sim$1--1.4       & \cite{Becker95, Scott03, Janka04, Miller15, Miller21}\\
$P$\,(ms)		& 33.4			& ATNF $^b$\\
$\dot P\,$ (s\,s$^{-1}$)		& $4.2\times10^{-13}$	& ATNF\\
$\Omega\,$(rad\,s$^{-1}$)		& 188.2			& ATNF\\
$\dot\Omega\,$(rad\,s$^{-2}$)			& $ -2.4\times 10^{-9}$	 & ATNF\\
$B_{ch}\,$(G)		& $3.8\times10^{12}$		& ATNF\\
$\tau\,$(yr)		& 1260		& ATNF\\
$\dot E \,$(erg\,s$^{-1}$)   & $4.5\times10^{38}$	& ATNF\\
$ L_{f} \,$(erg\,s$^{-1}$)   & $1.3\times10^{38}$	& \cite{Hester08}\\
\bottomrule
\end{tabularx}
\footnotesize{{\bf Note:} $^a$ The current parameters of the Crab pulsar based on the MDR, e.g., $\dot E$ and $\tau$. $^b$ ATNF Pulsar Catalogue \cite{Manchester05}: \url{https://www.atnf.csiro.au/research/pulsar/psrcat/} {(accessed on 1 March 2022)}.}
% \begin{flushleft}
% {\bf Note: } $^a$ The current parameters of the Crab pulsar based on the MDR, eg. $\dot E$ and $\tau$. $^b$ ATNF Pulsar Catalogue \cite{Manchester05}: https://www.atnf.csiro.au/research/pulsar/psrcat/
% \end{flushleft}
\end{table}

\section[\appendixname~\thesection]{Derivation of MDRW Model} \label{appendix d}
\renewcommand{\theequation}{A\arabic{equation}}
\setcounter{equation}{0}
%\newpage
%\section*{Appendix D: }\label{appendix d}
If we assume that the energy loss rate ($\dot E$) of pulsars is not only dipole radiation ($L_d$),
but may also contain other radiations, such as particle flow wind radiation ($L_f$), then we~obtain
\begin{equation}
\begin{split}
\dot E = L_d + L_f.
\end{split}
\end{equation}
The above equation can be written as
\begin{equation}
\begin{split}
I\Omega \dot \Omega = -(K_1 \Omega^4 + K_2 \Omega^2),
\end{split}
\end{equation}
where $K_1 = {2B^2R^6}/(3c^3)$ and $K_2=\eta B$ are parameters described in Sections~\ref{1} and \ref{2}; $I$ is the rotational inertia of a pulsar; $\Omega$ and $\dot \Omega$ are the angular velocity and its first derivative, respectively; $B$ is the real magnetic field strength of a pulsar; $R$ is the pulsar's radius; $c$ is the speed of light; and $\eta$ is the coefficient of the particle flow radiation.
Let $a=K_1/I$ and $b=K_2/I$, then the above equation can be simplified as
\begin{equation}
\begin{split}
\dot \Omega = -(K_1 \Omega^3 + K_2 \Omega)/I = -a\Omega^3 - b\Omega,
\end{split}
\end{equation}
and take the derivative of both
\begin{equation}
\begin{split}
\ddot \Omega = -(3a\Omega^2 + b)\dot \Omega.
\end{split}
\end{equation}
Two sides multiplied by $I\Omega^2$ can be obtained
\begin{equation}
\begin{split}
I\ddot \Omega \Omega^2 = -I(3a \Omega^4 + b \Omega^2 )\dot \Omega = (3L_d + L_f )\dot \Omega.
\end{split}
\end{equation}
The expression for the breaking index is
\begin{equation}
\begin{split}
n = \frac{\Omega \ddot \Omega}{\dot \Omega^2}.
\end{split}
\end{equation}
The numerator and denominator are multiplied by $I\Omega$ and take the Equation (A5) in it
\begin{equation}
\begin{split}
n = \frac{I\ddot\Omega\Omega^2}{(I\Omega\dot\Omega
)\dot \Omega}=\frac{(3L_d+L_f)\dot\Omega}{\dot E\dot \Omega}.
\end{split}
\end{equation}
We can simplify Equation (A1) to {obtain} %MDPI: please confirm if this equation is correct
\begin{equation}
\begin{split}
d + f = 1
\end{split}
\end{equation}
where $d=L_d/\dot E$ and $f=L_f/\dot E$ and Equation (A7) becomes
\begin{equation}
\begin{split}
n = 3 d + f.
\end{split}
\end{equation}
Therefore, solving Equations (A8) and (A9) simultaneously can be obtained
\begin{equation}
\begin{cases}
d=L_d/\dot E=(n-1)/2\\
f=L_f/\dot E=(3-n)/2.\\
\end{cases}
\end{equation}
Then, we determine the coefficients $K_1$ and $K_2$ by applying the present observed values  denoted by ``o'' for the Crab pulsar (B0531+21) breaking index ($n_{o}=2.50$), the energy loss rate ($\dot E_{o}=4.5\times10^{38}$\,erg\,s$^{-1}$), and angular velocity ($\Omega_o=188.2$\,rad\,s$^{-1}$)
\begin{equation}
\begin{cases}
L_{do}=(3/4)\dot{E_o}=K_1\Omega_o^4 \\
L_{fo}=(1/4)\dot{E_o}=K_2\Omega_o^2.\\
\end{cases}
\end{equation}
We take $I=10^{45}$\,g\,cm$^2$, $R=10^{6}$\,cm, and $c=3\times10^{10}$ cm\,s$^{-1}$ into calculation.
Thus, \mbox{$K_1=2.7\times10^{29}$\,c.g.s} and $K_2=3.2\times10^{33}$\,c.g.s, and $a=2.7\times10^{-16}$\,c.g.s and \linebreak \mbox{$b=3.2\times10^{-12}$\,c.g.s.}
%Therefore, we settle the parameters $B=3.3\times10^{12}\,G$ and $\eta=9.6\times10^{20}$.

Moreover, we know that $d$ and $f$ represent the proportion of $L_d$ and $L_f$ in $\dot{E}$, respectively.
Thus, the ratio factor $\epsilon$ between $f$ and $d$ is defined as (based on the definition of $\epsilon$, as can also~be seen in \cite{Lyne15})
\begin{equation}
\begin{split}
\epsilon = \frac{f}{ d} = \frac{3-n}{n-1} = \frac{b}{a}\Omega^{-2}=(\Omega_m/\Omega)^2=(P/P_m)^2,
\end{split}
\end{equation}
where $\Omega_m=108.6$\,rad\,s$^{-1}$ and $P_m=57.8$\,ms are the mean values of $\Omega$ and $P$ that correspond to $n=2$, respectively.

From Equations (A1) and (A2), we have two components of energy loss rate ($\dot{E}$).
We can simplify and solve the above differential equation and obtain the relationship between $\Omega$ and $t$.
\begin{equation}
\begin{split}
-\frac{d\Omega^2}{dt}=2(a\Omega^4+b\Omega^2),
\end{split}
\end{equation}
Here, we write $X = \Omega^2$ and the above equation can be rewritten as
\begin{equation}
\begin{split}
-\frac{dX}{dt}=2(aX^2+bX).
\end{split}
\end{equation}
Solve the above differential equation
\begin{equation}
\begin{split}
-\int\frac{1}{2(aX^2+bX)}dX=\int dt.
\end{split}
\end{equation}
Calculate the integral on both sides
\begin{equation}
\begin{split}
\frac{ln|a+b/X|}{2b}=t+C.
\end{split}
\end{equation}
Therefore, we obtain
\begin{equation}
\begin{split}
X=\Omega^2=\frac{b}{Ce^{2bt}-a}.
\end{split}
\end{equation}
We take the Crab pulsar's initial angular velocity ($\Omega_i=343.8$\,rad\,s$^{-1}$) and $t_i = 0$  yr into the above equation.
Then, constant $C = (\frac{b}{X_i}+a)$ is obtained by us, which can give the observed angular velocity value of the Crab pulsar at t = 960 years, implying that the exact solution is consistent.

Then, we can examine our calculation with the conditions of $b=0$ and $a=0$, respectively.
If $b=0$, then energy loss rate goes back to the MDR ($\dot{E} = L_d$), and we obtain the solution~below
\begin{equation}
\begin{split}
X_a=\Omega^2_a=\frac{1}{2at+1/X_i}.
\end{split}
\end{equation}
Meanwhile,
\begin{equation}
\begin{split}
X_{a1}&=\lim\limits_{b\rightarrow 0}\frac{b}{Ce^{2bt}-a}\\
&=\lim\limits_{b\rightarrow 0}\frac{b}{(b/X_i+a)(1+2bt)-a}\\
&=\frac{1}{2at+1/X_i}\sim X_a.
\end{split}
\end{equation}
On the other hand, if $a=0$, then the energy loss rate is occupied by particle flow radiation ($\dot{E} = L_f$). Thus, the solution of the equation can be obtained, that is
\begin{equation}
\begin{split}
X_b=\Omega^2_b=X_i e^{-2bt}.
\end{split}
\end{equation}
Correspondingly, Equation (A17) can be written as
\begin{equation}
\begin{split}
X_{b1}&=\lim\limits_{a\rightarrow 0}\frac{b}{Ce^{2bt}-a}\\
&=X_ie^{-2bt}\sim X_b.
\end{split}
\end{equation}
Then, if we take Equation (A17) into Equation (A12), we can obtain
\begin{equation}
\begin{split}
\epsilon=\frac{C}{a} e^{2bt}-1.
\end{split}
\end{equation}
Thus, Equation (A17) can be understood as a general solution of $\Omega^2$, and Equations (A18) and (A20) can be understood as special solutions.
Meanwhile, considering $P = 2\pi/\Omega$, we can obtain the relation between the spin period (P) and time (t).

When $L_d^{\prime}/\dot{E}=1$, $L_d^{\prime}=-{2B_{ch}^2R^6\Omega^4}/(3c^3)$, $B_{ch}=3.2\times10^{19}\sqrt{P\dot{P}}$ is the characteristic magnetic field strength.
We take $\dot{E}=L_d^{\prime}$ into $L_d/\dot{E}=(n-1)/2$, where $L_d$ is written by the real $B$, and $\dot{E}$ is written by $P$ and $\dot{P}$, and accordingly $B_{ch}$, so we have the relation between $B$ and~$B_{ch}$
\begin{equation}
\begin{split}
B^2/B_{ch}^2=(n-1)/2.
\end{split}
\end{equation}
Then, the evolution of $B$ can be written as
\begin{equation}
\begin{split}
B_{ch} = \sqrt{1+\epsilon}B = \sqrt{1+(P/P_m)^2}B.
\end{split}
\end{equation}

As Equation (A9) told us, breaking index (n) can be written as
\begin{equation}
\begin{split}
n = 2 d+1,
\end{split}
\end{equation}
and we take $ d = L_d/\dot{E}$ into it
\begin{equation}
\begin{split}
n = \frac{2L_d}{\dot E}+1=\frac{3+L_f/L_d}{1+L_f/L_d}=\frac{3+b/a\Omega^{-2}}{1+b/a\Omega^{-2}}.
\end{split}
\end{equation}
Furthermore, taking Equation (A12) into the above equation, we can obtain the evolution of the braking index (n) with the angular velocity ($\Omega$), that is
\begin{equation}
\begin{split}
n =\frac{3+\epsilon}{1+\epsilon} = \frac{2}{1+(P/P_m)^2}+1.
\end{split}
\end{equation}

The characteristic age is defined as $\tau_c=-\Omega/\dot{\Omega}$.
Thus, $\tau_c$ can be rewritten as
\begin{equation}
\begin{split}
\tau_c = -\frac{\Omega}{\dot{\Omega}} = -\frac{bI\Omega^2}{bI\Omega\dot{\Omega}} = -\frac{bI\Omega^2}{b\dot{E}} = \frac{L_f}{b\dot{E}}.
\end{split}
\end{equation}
Then, we take $L_f/\dot{E} = (3-n)/2$ into the above equation, and we can obtain
\begin{equation}
\begin{split}
\tau_c = \frac{1}{b} \frac{3-n}{2} = \frac{1}{b}\left(1-\frac{n-1}{2}\right).
\end{split}
\end{equation}
By Equation (A27), we have
\begin{equation}
\begin{split}
\tau_c &= \frac{1}{b}\left[1-\left(\frac{B}{B_{ch}}\right)^2\right] \\
&= \frac{\epsilon}{b(1+\epsilon)}\\
&= \frac{(P/P_m)^2}{b[1+(P/P_m)^2]}.
\end{split}
\end{equation}
Therefore, $\tau_{cmax} = 1/b = 3.15\times10^{11}\,s \approx 10^4\,yr$.

\section[\appendixname~\thesection]{More Explanation on MDRW Model} \label{appendix E}
%\section*{Appendix E: }
\label{appendix e}

Although the MDR model is simple, we rely on it to make a modification and study pulsar evolution by adding the wind flow component, and we would like to explain it in some detail.
To begin with, the MDR has been widely used in pulsar astronomy since it was created and it can describe the basic phenomena of pulsars.
Then, many parameters are calculated under MDR, such as the characteristic magnetic field and energy loss rate, which were recorded in the mature database of the ATNF Pulsar Catalogue  \cite{Manchester05}, as mentioned in the above text.
Meanwhile, it is a mainstream method to revise MDR by considering magnetosphere and wind, which has been discussed by many researchers \cite{Philippov22}.
For example, researchers have analyzed the wind on pulsar spin-down \cite{Kirk09, P19}, but they did not present the exact solutions for the Crab pulsar evolutions.
What we achieved is that the exact solutions for the Crab pulsar evolutions of spin, characteristic magnetic field, and braking index are obtained, by which we can quantitatively study the Crab pulsar compared with its well-measured observational data.
%In addition, to clearly show the wind flow in the pulsar radiation, we add an illustration figure of a pulsar magnetosphere.
Therefore, if MDR is totally discarded, this means that the parameters in the pulsar database will not be available for model calibration, and we keep the basic MDR and add the wind component in our MDRW model.

Furthermore, the vacuum environment assumed in the MDR is another widely concerning issue \cite{Shibata91, Cheng09, Kirk09, Li12, Petri16, Melrose16}, but we do not take this into account in the current work because the plasma propagation effect in the magnetosphere is quite intricate, which will let us lose the analytical expression of the MDRW.
Additionally, we do not consider the magnetic quadrupolar radiation in the present work,
%which will be considered in the subsequent research,
since the quadrupolar power should be related to the pulsar spin frequency in 5 power \cite{P19}, while for most pulsars, the braking index is lower than 3, which is not consistent with the observed value of the Crab pulsar---namely 2.5.
%the situations of the multi-polar magnetic field are worth to be further analysed.
However, the magnetic quadrupolar structure is expected to be complicated \cite{Krolik91, P12, Petri15, P17, P20a, P20b}, which is far beyond our scope.
%that we would spend some time to handle its formation mechanism and consider how to apply it to the Crab pulsar.
Thus, in the present state, the MDRW is a concise and focused model to explain the braking index and evolutions of the Crab pulsar.
%, and in the future the extended MDRW that considers the above conditions will be more suitable for many other physical evolution paths of pulsars.

In the end, to avoid potential confusion, the issue of magnetars from the Crab pulsar should be clarified, because the simple MDRW model indicates that the characteristic B-field of the Crab pulsar grow from $\sim$10$^{12}$\,G to $\sim$10$^{14}$\,G, while the true B-field of the Crab pulsar has no change at all, which possibly hints at the potential existence of the  ``impostor'' magnetars. 
However, this association is ambiguous and unrealistic, the reasons for which are considered below.
To begin with, the conceptions of magnetars are based on soft gamma-ray repeaters (SGRs) and anomalous X-ray pulsars (AXPs) that often exhibit the intense high-energy outbursts, which are well-explained  by the assumed super-strong true B-field magnetars \cite{Duncan92,Thompson93, Ferrario08, Kaspi17, Esposito21}.
However, the outburst phenomena of SGRs/AXPs are also considered as other interactions between the wind flow  materials or fall-back disk and the magnetosphere \cite{Alpar01, popov06, Bisnovatyi-Kogan17}.
Then, the emission properties of the Crab pulsar are significantly different from those of the magnetars, and the persistent  X-ray luminosity ($L_x$)  of the magnetar is often higher than its spin-down kinetic energy loss rate, inferring that both sources should have different origins and properties \cite{Mereghetti15, Kaspi17, Esposito21}.
{It is remarked that perhaps the true B-field of some magnetars may be overestimated because the electron X-ray absorption cyclotron line has not yet measured them} \cite{Truemper78, Ye19}---which is now also estimated and inferred by the spin period $P$ and its derivative based on the MDR model.
Conservatively and possibly, MDRW may be applied to several high characteristic B-field radio pulsars such as PSR J1734-3333 \cite{Espinoza11}; meanwhile, it shows a low braking index close to 1, which is consistent with the prediction of the wind flow torque itself. 
Thus, if the MDRW model dominates its spin-down torque, we just guess and assume its true B-field to be approximately $\sim$10$^{12}$\,G. 
Nevertheless, we cannot assure this conclusion, since there is no direct measurement of the true B-field for this source.
Additionally, the Galactic magnetar SGR \mbox{1935 + 2154} generated fast radio bursts (FRBs) \cite{CHIME20, Bochenek20}, supporting the hypothesis of the magnetar origin of FRBs\cite{Popov10, Cui22}.
However, the properties of FRBs are far from  the  pulses of normal radio pulsars\cite{Petroff22}, such as energy, polarization, and narrow band,  indicating that some basic differences between both  should  exist.
In other words, we stress that it is too early to apply the simple  MDRW model to all types of NSs, because their origins and formation mechanisms may be different \cite{popov06,Kaspi10, Beskin18, Cui21, Zhang22}.

\begin{adjustwidth}{-\extralength}{0cm}
%\centering %% If there is a figure in wide page, please release command \centering

\reftitle{References}

% Please provide either the correct journal abbreviation (e.g., according to the “List of Title Word Abbreviations” http://www.issn.org/services/online-services/access-to-the-ltwa/) or the full name of the journal.

%\bibliography{your_external_BibTeX_file}

%=====================================
% References, variant B: internal bibliography
%=====================================

\end{adjustwidth}
\end{document}